\documentclass[a4paper,12pt]{article}
\usepackage[utf8]{inputenc}
\usepackage{amsmath}
\usepackage{amsfonts}
\usepackage{amssymb}
\usepackage{graphicx}
\usepackage{hyperref}
\usepackage{authblk}

\begin{document}

\title{Desertification by Front Propagation?}

\author[1]{Yuval R. Zelnik}
\author[2]{Hannes Uecker}
\author[3]{Ulrike Feudel}
\author[1,4]{Ehud Meron}

\affil[1]{Department of Solar Energy and Environmental Physics, Blaustein Institutes for Desert Research, Ben-Gurion University of the Negev, Sede Boqer Campus, 84990, Israel}
\affil[2]{Institute for Mathematics, Carl von Ossietzky University Oldenburg, P.F 2503, 26111 Oldenburg, Germany}
\affil[3]{Institute for Chemistry and Biology of the Marine Environment, Carl von Ossietzky University Oldenburg, PF 2503, 26111 Oldenburg, Germany}
\affil[4]{Department of Physics, Ben-Gurion University of the Negev, Beer Sheva, 84105, Israel}

\maketitle

\begin{abstract}
Understanding how desertification takes place in different ecosystems is an important step in attempting to forecast and prevent such transitions.
Dryland ecosystems often exhibit patchy vegetation, which has been shown to be an important factor on the possible regime shifts that occur in arid regions in several model studies.
In particular, both gradual shifts that occur by front propagation, and abrupt shifts where patches of vegetation vanish at once, are a possibility in dryland ecosystems due to their emergent spatial heterogeneity.
However, recent theoretical work has suggested that the final step of desertification - the transition from spotted vegetation to bare soil -
occurs only as an abrupt shift, but the generality of this result, and its underlying origin, remain unclear.
We investigate two models that detail the dynamics of dryland vegetation using a markedly different functional structure, and find that in both models the final step of desertification can only be abrupt.
Using a careful numerical analysis, we show that this behavior is associated with the disappearance of confined spot-pattern domains as stationary states, and identify the mathematical origin of this behavior.
Our findings show that a gradual desertification to bare soil due to a front propagation process can not occur in these and similar models, and opens the question of whether these dynamics can take place in nature. 
\end{abstract}

\section{Introduction}
Desertification is a major concern in water-limited ecosystems or drylands, which occupy about $40\%$ of the terrestrial earth surface.
It is defined as a transition from a productive state to a less productive state as a result of climate variability and human-induced disturbances \cite{maweb-2005-desertification,Meron2015book}.
Drylands are prone to such transitions because they can often assume two alternative stable states of vegetation.
Bistability of vegetation states is induced by positive feedbacks involving various biotic and abiotic processes \cite{Rietkerk2004science,Meron2012eco_mod,Meron2016mb}.
A productive vegetation state is stabilized by enhanced surface-water infiltration,
reduced evaporation, litter decomposition that increases nutrient availability,
soil deposition and mound formation that intercept runoff, etc. Under the same environmental conditions the unproductive bare soil state is stabilized by enhanced evaporation,
water loss by runoff that is generated by soil crusts, soil erosion, etc.

A significant aspect of dryland ecosystems is that they often self-organize in regular spatial patterns of vegetation in response to decreasing rainfall \cite{Valentin1999catena,Deblauwe2008geb,Rietkerk2008tree,getzin2016discovery}.
This is a landscape-level mechanism to cope with water deficit by providing an additional source of water through various means of water transport.
The mechanism involves positive feedbacks between local vegetation growth and water transport towards the growth location \cite{Meron2016mb}.
While accelerating the growth of a vegetation patch, the water transport inhibits the growth in the patch surroundings and thereby favors the formation of spatial patterns \cite{lejeune-1999-short,Rietkerk2004science,Gilad2004prl}.
Among the water transport mechanisms that have been identified are overland water flow, soil-water diffusion and water conduction by laterally extended root systems \cite{Meron2016mb}.
The capability of these and additional feedbacks to destabilize uniform vegetation and produce patterns as precipitation drops below a threshold value has been verified in many model studies \cite{Lefever1997bmb,Klausmeier1999science,vonHardenberg2001prl,HilleRisLambers2001ecology,Gilad2004prl,sherratt-2005-analysis,vanderStelt2012nonl_sci}.

As rainfall further decreases, the water-contributing bare soil areas need to increase in order to compensate for the lower rainfall the vegetation patches directly receive.
The increase of bare soil area can occur in three distinct ways: (i) contraction of vegetation patches, keeping the pattern's wavenumber constant \cite{Yizhaq2005physicaA},
(ii) transitions to periodic patterns of lower wavenumbers, keeping the pattern morphology unchanged \cite{Siteur2014eco_comp},
(iii) morphology changes from patterns of gaps to stripes to spots \cite{vonHardenberg2001prl,Rietkerk2002an,Lejeune2004ijqc, Gilad2007jtb, Gowda2014pre, Gowda2016prsa}.

Another spatial aspect of dryland ecosystems that bears on desertification is the confinement of typical disturbances to relatively small area.
Rather than inducing a global shift to the alternative state, such disturbances can induce local shifts.
The subsequent time behavior depends on the dynamics of the fronts that connect the two alternative states, i.e. which state invades the other, on front interactions and possibly on front instabilities \cite{Hagberg1994nonlinearity,hagberg1994complex}.
An unproductive bare soil state invading a productive vegetation state, uniform or patterned, is a form of desertification taking place gradually by front propagation \cite{Bel2012theo_ecol}.
When one of the alternative stable states is spatially patterned the front may be pinned in place in a range of environmental conditions
\cite{Pomeau1986pd,Knobloch2015annurev-conmatphys} nested within the bistability range, the so-called ``snaking range''.
In this range a multitude of stable hybrid states exist in addition to the two alternative states.
These are spatially-mixed states consisting of confined domains of one state in a system otherwise occupied by the other state \cite{Kozyreff2006prl,burke2007homoclinic,Lloyd2008sjads}.
Depending on environmental variability, the effect of local disturbances in this case can remain local \cite{Bel2012theo_ecol}.

Understanding how a desertification process may occur, taking into account the possible effects of an alternative state being spatially patterned and the confined nature of disturbances,
is vital for both finding indicators for an impending desertification, and for efforts to its prevention.
The process of gradual desertification by front propagation has been first demonstrated using a minimal model \cite{Bel2012theo_ecol},
equivalent to the Swift-Hohenberg equation for which a snaking range nested within a bistability range is known to exist \cite{Knobloch2008nonlinearity}.
In the context of dryland vegetation, this behavior suggests the existence of confined domains of patchy (patterned) vegetation in a bare soil state within the snaking range,
which are fixed in size, i.e. neither expanding nor contracting, because of front pinning,
and the existence of a ``desertification front'' (bare soil invading patchy vegetation), and a ``recovery front'' (patchy vegetation invading bare soil) outside that range, on the lower and higher rainfall sides, respectively.

In a subsequent study Zelnik et al. \cite{Zelnik2013regime} have investigated how these ideas translate when considering more ecologically motivated models.
Looking at different models of dryland ecosystems which exhibit spatial patterns of vegetation,
they were unable to find localized states describing a confined patchy-vegetation domain in bare-soil in most models \cite{Zelnik2013regime}.
One model investigated, herewith referred to as the Lefever-Lejeune (LL) model \cite{Lejeune2004ijqc}, did appear to show localized states, although this was not definitively shown due to numerical difficulties.
Perhaps most strikingly, in all models considered no bistability range where a bare-soil domain invades the patchy vegetation was seen \footnote{Note that bare-soil can invade a spatially uniform-vegetation state \cite{Sherratt2012vegetation}.}.
This is countered with the dominant response of the system to different perturbations, where fronts of patchy vegetation takes over the system,
due to vegetation on the fringe utilizing the resources in the bare-soil domain to its advantage as it expands into it.

The reason for this asymmetry between desertification and recovery transitions has not been explained, and it remains unclear how general these results are.
Moreover, since the typical behavior of hybrid states within a snaking range, and desertification and recovery fronts outside this range,
can be seen in the same models but in a bistability range of patchy vegetation and uniform-vegetation (rather than bare-soil) \cite{zelnik2015pnas},
understanding the unique characteristics of the bare-soil and patchy-vegetation bistability is an important ecological question.
In this paper we consider two models of dryland ecosystems focusing on the bistability range of bare-soil and periodic vegetation.
The first is the LL model while the second is a simplified version of a model introduced by Gilad et al. \cite{Gilad2004prl,Gilad2007jtb},
hereafter referred to as the simplified Gilad (SG) model. We will investigate the existence and structure of localized states in these models, and look at the dynamics in their vicinity.
In particular, we use the flexibility of the SG model to show how the bifurcation structure of localized states breaks down with a continuous change in parameter values,
give evidence that this breakdown is related to the spatial eigenvalues of the bare-soil and low uniform-vegetation states,
and that thus is a generic phenomenon in the class of reaction diffusion models for drylands.

\section{Bistability, fronts and homoclinic snaking}
We begin with a brief review of front dynamics in one-dimensional spatially extended bistable systems, according to pattern formation theory \cite{Meron2015book}. 
There are two major types of bistability that are relevant to dryland vegetation, bistability of two uniform states and bistability of a uniform state and a periodic-pattern state. 
The former can be found in landscapes where positive feedbacks that involve water transport, and thus are pattern forming, are too weak to destabilize uniform vegetation, 
but other positive feedbacks, such as reduced evaporation in vegetation patches, are sufficiently strong. 
These conditions can result in bistability of bare soil and uniform vegetation \cite{getzin2016discovery}. 
Bistability of uniform and patterned states can result when the pattern-forming feedbacks that involve water transport are strong enough to induce a subcritical instability of uniform vegetation to periodic patterns. 
Two bistability precipitation ranges of uniform and patterned states can be distinguished: a high precipitation range where the two alternative stable states are uniform vegetation and periodic patterns, 
and low precipitation range where the two alternative states are periodic patterns and bare soil\cite{Meron2016mb}.

\begin{figure}[h]
 \includegraphics[width=0.99\textwidth]{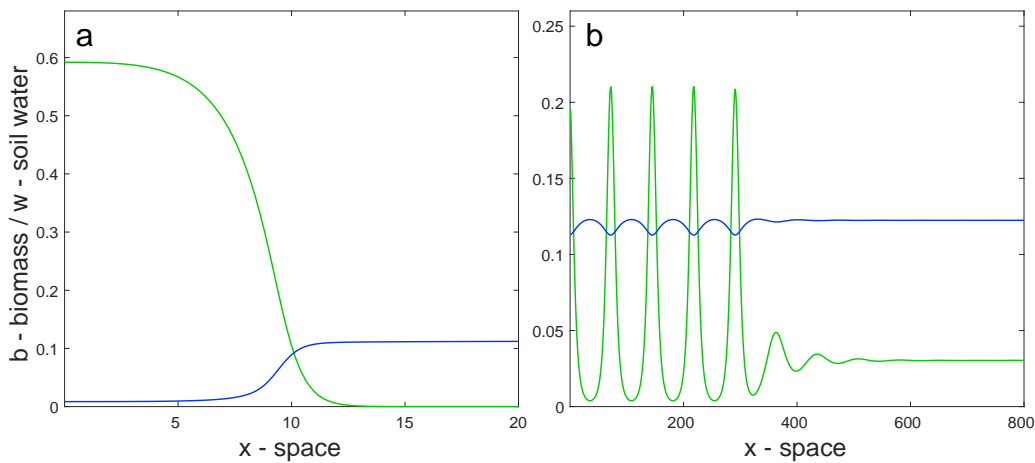}
 \caption{ \footnotesize Spatial profiles of two types of fronts in the SG model\protect\footnotemark[2].
 (a) A front between uniform-vegetation and bare-soil. 
 (b) A front between periodic patterned vegetation and uniform-vegetation.
 Biomass and soil-water profiles are shown in green and blue, respectively.
  \label{fig:Fronts} }
\end{figure}

When different parts of the landscape are occupied by different states, transition zones appear, where some of the state variables, e.g. biomass, sharply change. 
Such zones, often called "fronts", have characteristic structures determined by particular spatial profiles of the state variables across the front, as Fig. \ref{fig:Fronts} illustrates.

\footnotetext[2]{See the Models section for a description of the SG model. The parameters used were based on (\ref{sgpar}), with the addition of: (a) $\delta_w=1200$, $p=1.008$, (b) $\delta_w=1$, $p=0.9$, $\eta=6$.}

Furthermore, the fronts can be stationary or moving at constant speeds (assuming a homogeneous system). In a bistability range of two uniform states
fronts are generically moving. The speed and direction of front motion are determined by the values of various parameters that describe specific biotic and abiotic conditions, or by the specific value of a control parameter, where all other parameters are held fixed. A particular control-parameter value for which the front is stationary may exist, but this is a non-generic behavior as any deviation from this value results in front motion \cite{Bel2012theo_ecol}. In a bistabilty precipitation range of uniform vegetation and bare soil two moving fronts can be distinguished; a desertification front at low precipitation that describes the expansion of bare-soil domains into uniform-vegetation domains, and a recovery front at high precipitation that describes the expansion of uniform-vegetation domains into bare-soil domains \cite{Sherratt2012vegetation}. The particular precipitation value at which the non-generic behavior of a stationary front occurs (neither domain expands into the other) is commonly called the ``Maxwell point''.

In contrast to bistability of uniform states, when one of the alternative states is a periodic pattern, fronts can be stationary or pinned in a \emph{range} of the control parameter \cite{Pomeau1986pd,Knobloch2008nonlinearity}. 
In this range domains of one state embedded in the alternative state can remain fixed in size, neither expanding nor contracting.  
In a bifurcation diagram that describes the biomass associated with stationary solutions vs. precipitation, 
these fixed domains usually appear as a single or two branches of localized solutions that snake back and forth as the sizes of the domains they represent change (see e.g. Fig.  \ref{fig:SGdw1200}),
a behavior termed homoclinic snaking. The snaking solution branches occupy a subrange of the bistability range - the snaking range. 
Within this range a multitude of spatially localized and extended stable hybrid states exist, corresponding to single fixed domains and various combinations of such domains, respectively. 
Thus, in a bistability range of uniform and patterned states three front types can be distinguished, a stationary pinned front within the snaking range and two fronts moving in opposite directions on both sides of this range. 
In a bistability range of uniform vegetation and periodic vegetation patterns the two moving fronts represent a desertification front at precipitation values below the snaking range 
(a periodic pattern displacing uniform vegetation) and a recovery front at precipitation values above the snaking range (uniform vegetation displacing a periodic pattern).

These front properties become significant in the presence of local disturbances that induce local shifts to an alternative state, 
and rainfall fluctuations that can take the system outside the snaking range \cite{Bel2012theo_ecol,zelnik2015pnas}. 
Local disturbances can trigger gradual desertification at precipitation values below the Maxwell point in bistability of uniform vegetation and bare soil, and below the snaking range in bistability of uniform vegetation and periodic patterns. 
Within the snaking range, local disturbances have little effect as the system converges to a nearby hybrid state. 
Droughts that take the system temporarily out of the snaking range can induce transitions to hybrid states of lower productivity, but result in no further effects once the droughts are over. 
The relevance of these outcomes to the bistability range of periodic patterns and bare soil, however, is not clear. 
In most models homoclinic snaking has not been found, which excludes stationary pinned fronts and hybrid states. 
Moreover, only one type of moving fronts has been found - recovery fronts \cite{Zelnik2013regime}. 
The absence of desertification fronts suggests that desertification to bare soil occurs abruptly, by global vegetation collapse \cite{zelnik2016localized}.

\section{Models}
We study the LL and SG models assuming a flat terrain (no slope) and a homogeneous system (no environmental heterogeneities, such as rock-soil mosaics), and thus have both translation and reflection symmetries.
For simplicity we limit the results shown here to one-dimensional systems, but we exemplarily checked that similar results also hold in two spatial dimensions (2D)
(see supplementary information for more details), and thus believe our findings to be generally relevant in 2D as well.

The dimensionless LL model \cite{Lefever1997bmb,Lejeune2002pre}  is given by the single equation
\begin{equation}\label{eq:LL}
 \partial_t b = \left(p-1 \right)b+\left(\Lambda-1\right)b^2-b^3+\frac{1}{2}\left(L^2-b\right)\nabla^2 b-\frac{1}{8}b\nabla^4 b\,.
\end{equation}
where $b$ is the biomass density, $p=2-\mu$ where $\mu$ is the aridity parameter, $\Lambda$ is the facilitation to competition ratio,
and $L$ is the ratio between the spatial ranges of facilitation and competition interactions, and $\nabla^2$ is the Laplacian, i.e., in 1D, $\nabla^2=\partial_x^2$,
and similarly $\nabla^4=\partial_x^4$..
We will look at different values of $p$, while keeping the other two parameters constant with the values $\Lambda = 1.2$, $L=0.2$.

The SG model \cite{Gilad2007jtb} has the non-dimensional form
\begin{subequations}\label{eq:SG}
  \begin{align}
      \partial_t b & = \nu w b (1+\eta b)^2 (1-b/\kappa) - b + \nabla^2 b, \\
      \partial_t w & = \alpha \frac{b+q f}{b+q} h - \nu w(1-\rho b/\kappa) - \nu w b (1+\eta b)^2 + \delta_w \nabla^2 w, \\
      \partial_t h & = p - \alpha \frac{b+q f}{b+q} h + \delta_h \nabla^2 (h^2),
  \end{align}
\end{subequations}
where the three model variables describe the areal densities of vegetation biomass ($b$), soil-water ($w$), and surface water ($h$) \cite{Kinast2014prl}.
The parameter $\nu$ controls the growth rate of the biomass as well as the rate of evaporation, $\eta$ gives the root-to-shoot ratio, and $\kappa$ is the maximal standing biomass.
The parameter $\alpha$ controls the infiltration rate, while $f$ and $q$ quantify the biomass dependence of the infiltration rate.
The reduced evaporation in vegetation patches because of shading is quantified by $\rho$.
Finally, the parameter $p$ is the precipitation rate and $\delta_w$ and $\delta_h$ are water-transport coefficients associated with soil water and surface water, respectively.
The SG model is a simplified version of a more general model \cite{Gilad2004prl} that takes into account nonlocal water uptake by laterally extended root systems.
We refer the reader to refs. \cite{Meron2015book,Meron2016mb} for detailed expositions of the general model and its simplified versions.
We will look at different values of two parameters, $\delta_w$ and $p$, while keeping the others constant.  The values of these parameters are

\begin{gather}\label{sgpar}
\text{ $\nu = 8.0$, $\kappa = 2.0$, $\eta = 0.6$, $\alpha = 25.0$, $q = 0.70$, $f = 0.9$, $\rho = 0.15$, $\delta_h = 10000$.}
\end{gather}

The domain used for the state plots, and for calculating the norms of the bifurcation diagrams was 100/1000 for the LL/SG models, respectively.
The bifurcation structure for the LL and SG models was found using the numerical continuation tools of pde2path \cite{uecker2014pde2path} and AUTO \cite{auto}, respectively.
The stability information of these was found using numerical linear stability analysis on a large domain with periodic boundary conditions (size of approx. 100/1000 for the LL/SG models, in integer number of periods).
To clearly separate the different solution branches in the bifurcation diagrams of Fig. \ref{fig:SGdw1200} and Fig. \ref{fig:SGdw1500},
the y-axis for these were calculated using the L2norm of the biomass $b$ across a domain $L$: $||b|| = \sqrt{\frac{1}{L}\int_0^L dx \cdot (b(x))^2}$.
The y-axis for the bifurcation diagrams of Fig. \ref{fig:LLbif} was calculated using the average biomass $b$ across a domain $L$: $|b| = \frac{1}{L}\int_0^L dx \cdot b(x)$.

\section{Localized states in the LL model}
We begin by looking at the LL model, around the bistability range of periodic patterns and a uniform bare-soil state.
In this model, uniquely among models of dryland vegetation that exhibit pattern formation, localized states where the uniform background is of bare-soil can be found.
This may be attributed to the nonlinear degeneracy of the LL model at the bare-soil, which does not occur for the SG model since $h>0$ even when $b=0$ (see Fig. S4).

\begin{figure}[h]
\includegraphics[width=0.755\textwidth]{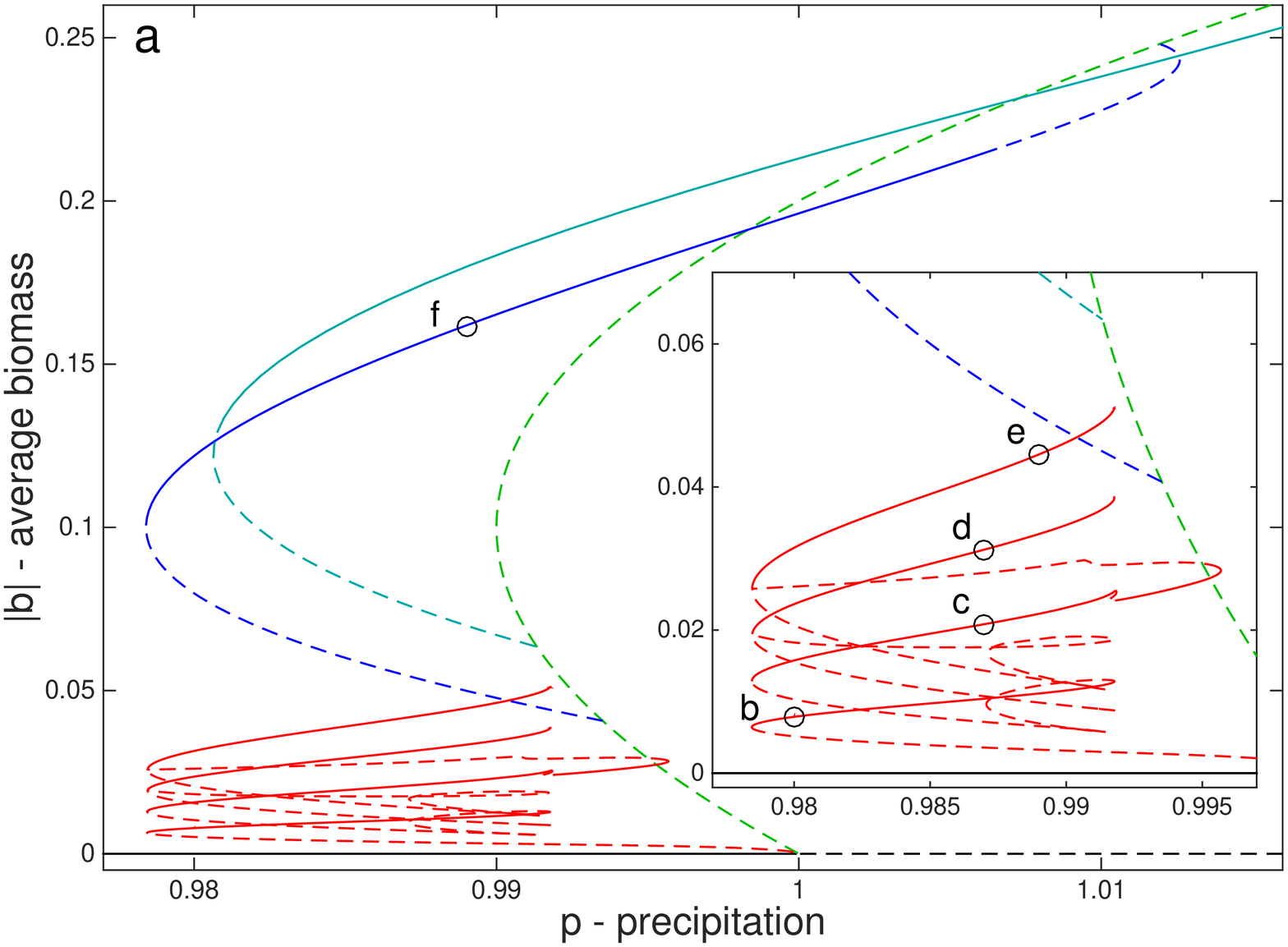}
\includegraphics[width=0.2355\textwidth]{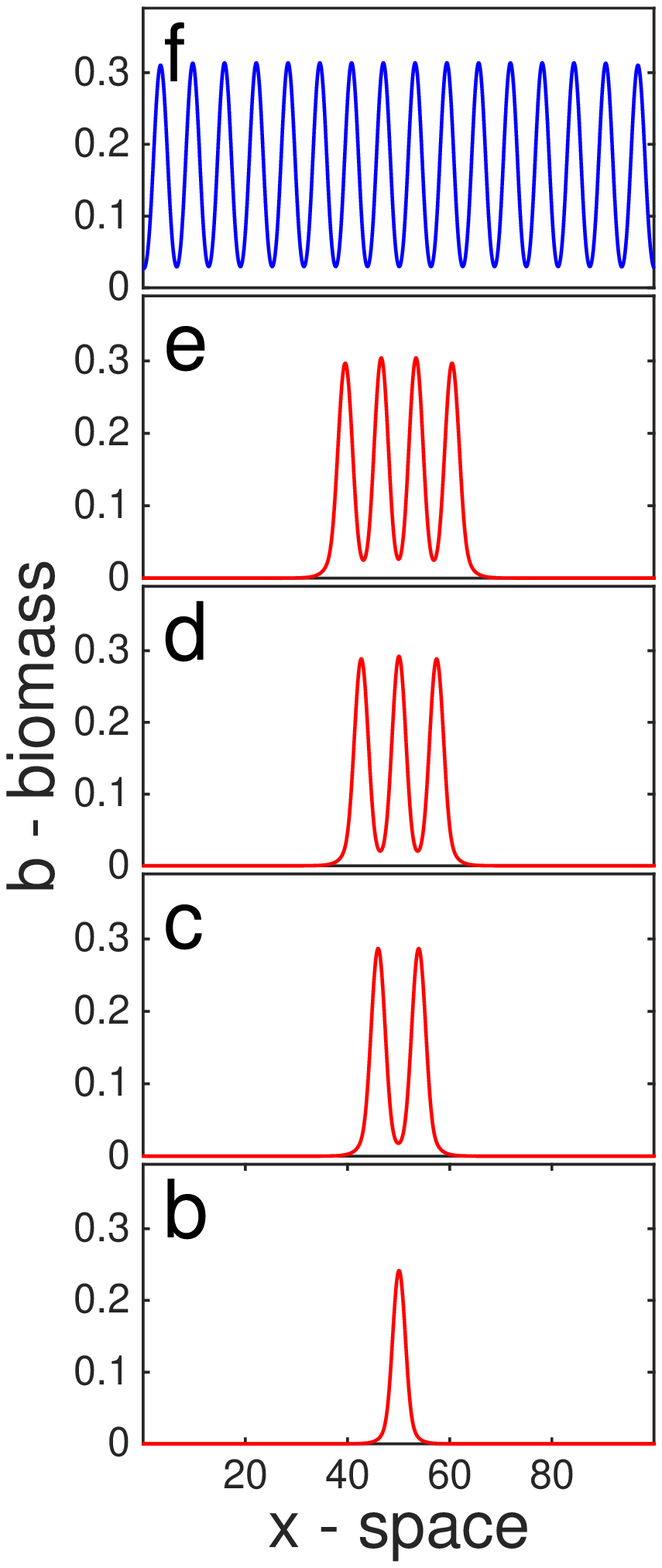}
 \caption{ \footnotesize (a) Bifurcation diagram of the LL model, showing the average biomass versus the bifurcation parameter $p$.
 The black and green branches are uniform states of bare-soil and uniform-vegetation, while the blue and cyan branches are of periodic states with different wavelength (many more exist).
 The red curve shows part of the intricate bifurcation structure of the localized states.
 The left-most point of the blue curve defines the edge of the bistability range between periodic states and bare-soil, and it coincides with the edge of the range of the localized states.
 Solid (dashed) lines denote stable (unstable) states.
 (b)-(f): Plots of one periodic state and four localized states, with their location on the bifurcation diagram denoted by their letters in the main diagram.
 \label{fig:LLbif} }
 \end{figure}

Inside the bistability range there are stable localized states in a subrange we refer to as the snaking range, as can be seen in Fig. \ref{fig:LLbif}.
While the numerics for the uniform and periodic branches in this bifurcation diagram are straightforward, the branches of localized states are difficult to generate and incomplete in the following sense.
The states (b),(c) are connected by a continuous branch while (d), (e) were obtained from restarting the continuation with different initial guesses,
and it is not entirely clear how their branches connect to (b),(c) and with each other.
Nevertheless, we refer to the red branches as the snaking branches, and the associated $p$--range as the snaking range.

\begin{figure}[h]
 \includegraphics[width=0.99\textwidth]{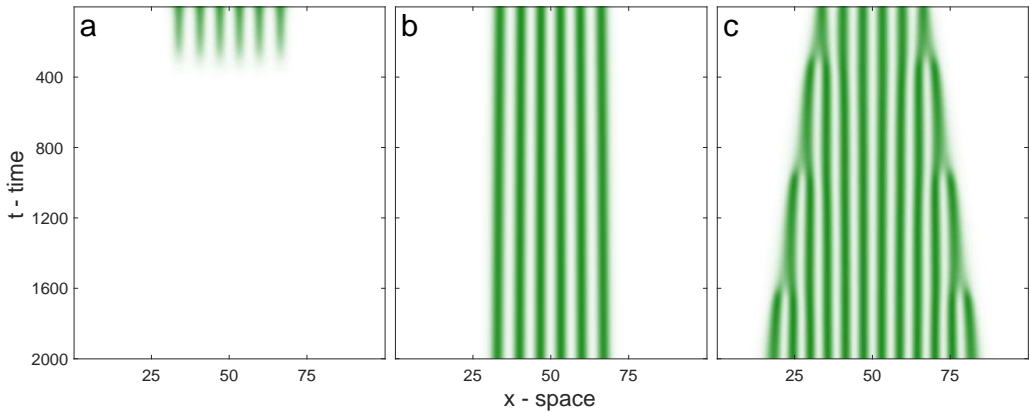}
 \caption{ \footnotesize Space-time plot showing the dynamics of the LL model, starting with an initial condition of a stable localized state with six peaks.
  Darker green colors show denser vegetation, with the y-axis for time (going down) and the x-axis for space.
 (a) The value $p$ is decreased, so that the system is taken out of the snaking range ($p=0.976$).
 Since this is also outside the bistability range, no gradual transition is possible, and an abrupt desertification shift occurs.
 (b) The value of $p$ is not changed, so that the system is still inside the snaking range ($p=0.990$), and therefore no change occurs.
 (c) By increasing the value of $p$ so it is outside the snaking range ($p=0.998$), but still within the bistability range, a gradual rehabilitation process takes place.
  \label{fig:LLdyn} }
\end{figure}

Outside the snaking range, localized states do not exist, so that a transition from a localized state will occur if $p$ is changed significantly.
If we increase $p$ slightly, the system is still in a bistability range, so that a gradual transition by the propagation of a recovery front can occur, as shown in Fig. \ref{fig:LLdyn}c.
The recovery front propagates by the expansion of the outermost patch, which experiences reduced competition,
and the subsequent patch splitting, because of increased competition at the center of the growing patch \cite{Sheffer2007eco_comp}.
However, a change in the other direction, namely decreasing $p$ even only slightly outside the snaking range, will take the system out of the bistability range as well.
Since periodic states do not exist for these parameters, a gradual transition by front propagation cannot occur, and an abrupt one takes place instead, as seen in Fig. \ref{fig:LLdyn}a.

To our knowledge, this asymmetry where the snaking range is not in the middle of the bistability range has not been observed in other models.
While its ramifications for the dynamics in the system are clear, as shown in  Fig. \ref{fig:LLdyn}, the reason for this asymmetry is not understood.
To explore this question further, we will now use the flexibility of the SG model, and its somewhat simpler and more classical mathematical structure as a semi--linear reaction diffusion system.

\section{Localized states in the SG model}
A typical bifurcation structure for the SG model, much like other similar models (e.g. the Rietkerk model \cite{HilleRisLambers2001ecology}),
shows a bistability range between periodic patterns and a bare-soil state, but no localized states are found within this bistability range, except for a single-peak solution.

However, if we soften the criteria slightly, and look at a bistability between periodic patterns and low biomass uniform-vegetation,
localized states in a classic bifurcation structure of homoclinic snaking can be found \cite{dawes2015localised}.
This may occur when the bare-soil branch goes through a supercritical bifurcation to the uniform-vegetation branch, on which at some finite distance from this primary bifurcation there is a subcritical Turing bifurcation.
For the SG model, this can be achieved with a low shading feedback ($\rho \gtrsim 0$), a medium level of root-to-shoot ratio ($\eta \approx 1$), and fast water diffusion ($\delta_w \gg 1$)

\begin{figure}[h]
 \includegraphics[width=0.755\textwidth]{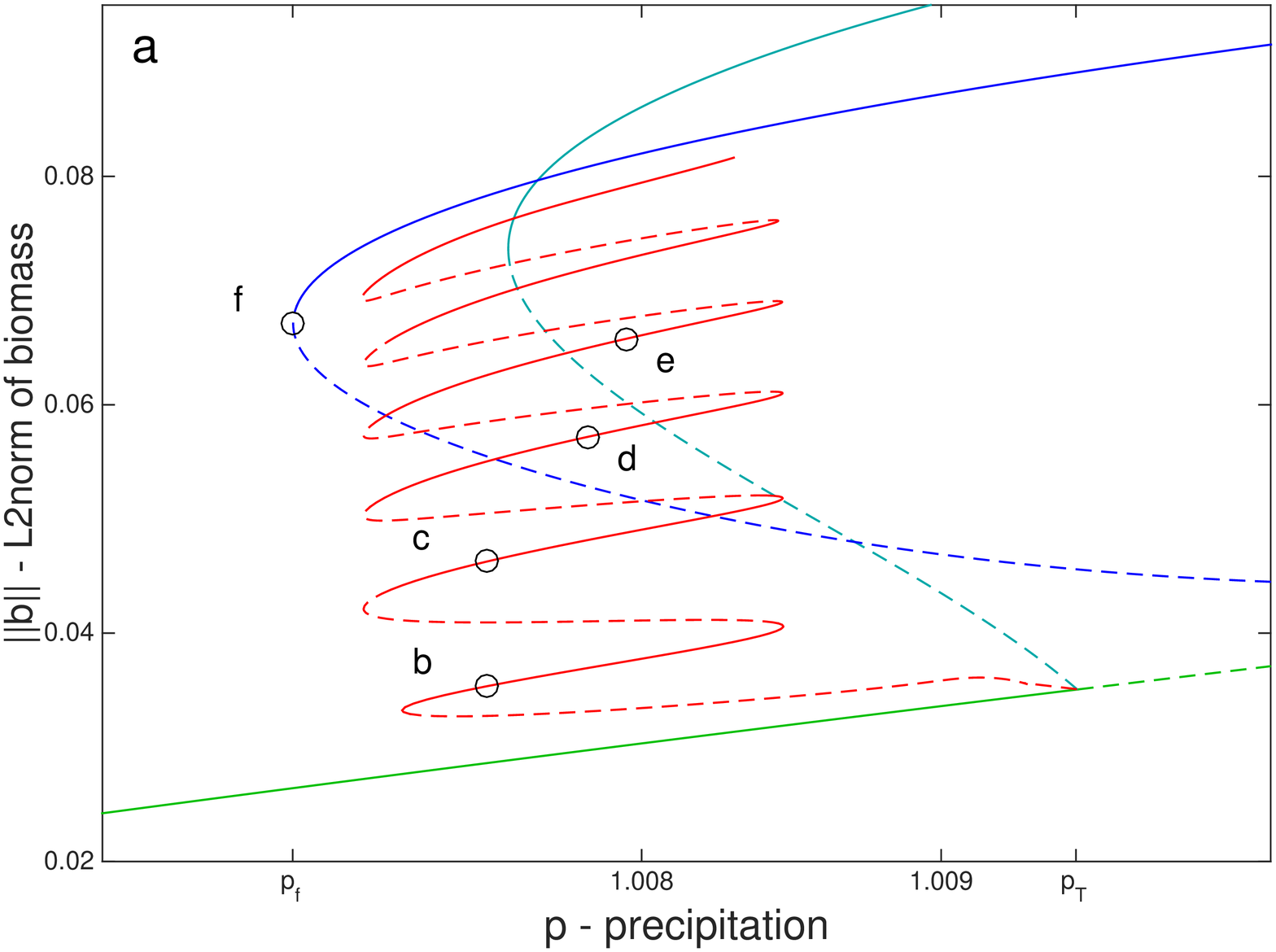}
 \includegraphics[width=0.2355\textwidth]{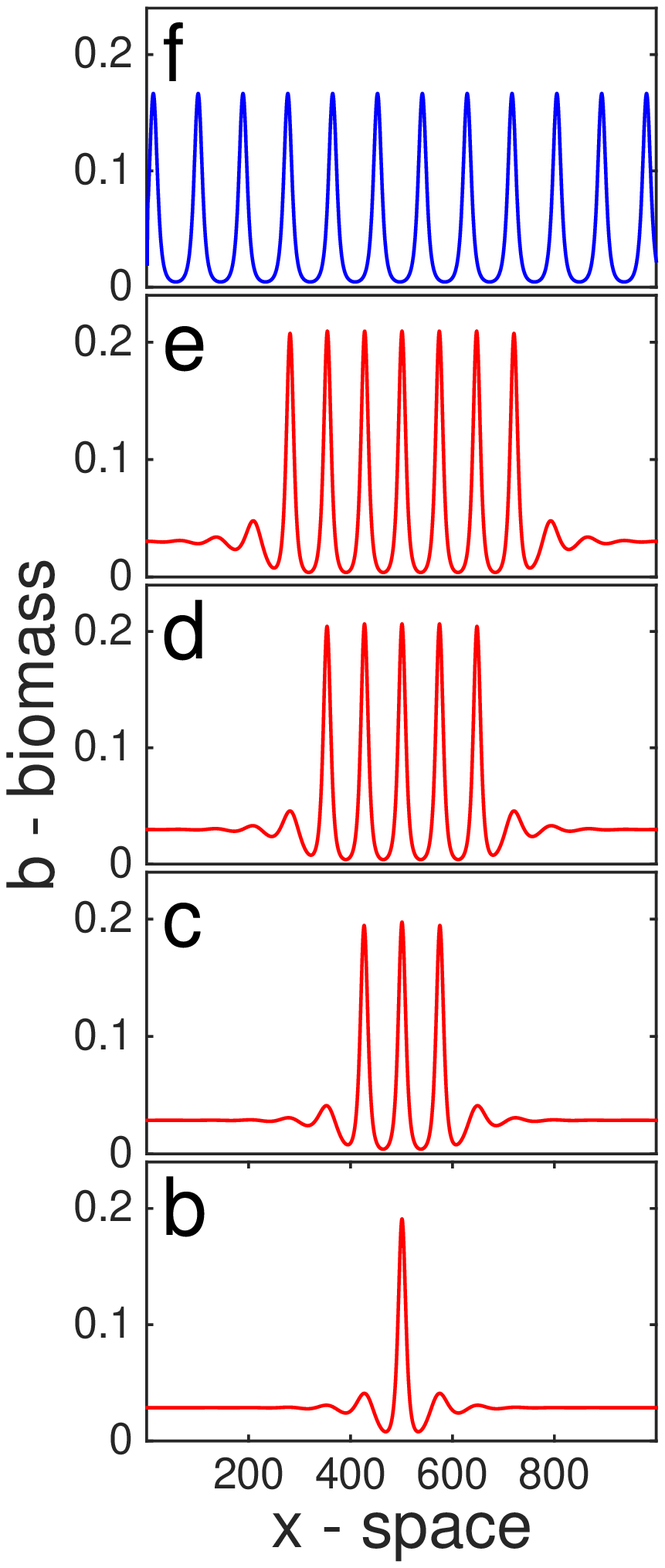}
 \caption{ \footnotesize
 (a) Bifurcation diagram of the SG model with parameters (\ref{sgpar}) and $\delta_w=1200$, showing the L2norm of the biomass versus the precipitation rate $p$.
 The green branch represents states of uniform-vegetation, while the blue and cyan branches are for periodic states with different wavelengths.
 The red curve shows part of the bifurcation structure of the localized states, termed homoclinic snaking.
 The wavelength of the blue curve was chosen so that its left-most point defines the edge of the bistability range between periodic states and low uniform-vegetation (see Fig. S1),
 and it is well away from the edge of the range of the localized states (thus fronts of low vegetation invading the periodic states are possible, cf. Fig. \ref{fig:SGdyn}b.)
 Solid (dashed) lines denote stable (unstable) states.
 (b)-(f): Plots of one periodic state and four localized states, with their location on the bifurcation diagram denoted by their letters in the main diagram.
\label{fig:SGdw1200}}
 \end{figure}

Using (\ref{sgpar}) and $\delta_w=1200$, we arrive at the bifurcation diagram shown in Fig. \ref{fig:SGdw1200}.
Emanating near the first subcritical Turing bifurcation on the supercritical uniform-vegetation branch there is a branch of localized states in a homoclinic snaking structure,
within a larger bistability range of periodic patterns and uniform-vegetation.
The branch of localized states is initially that of a single peak in a background of low uniform-vegetation (Fig. \ref{fig:SGdw1200}b).
As the branch snakes up more peaks are attached, forming a domain of semi-periodic vegetation (Fig. \ref{fig:SGdw1200}c-e).

\begin{figure}[h]
 \includegraphics[width=0.99\textwidth]{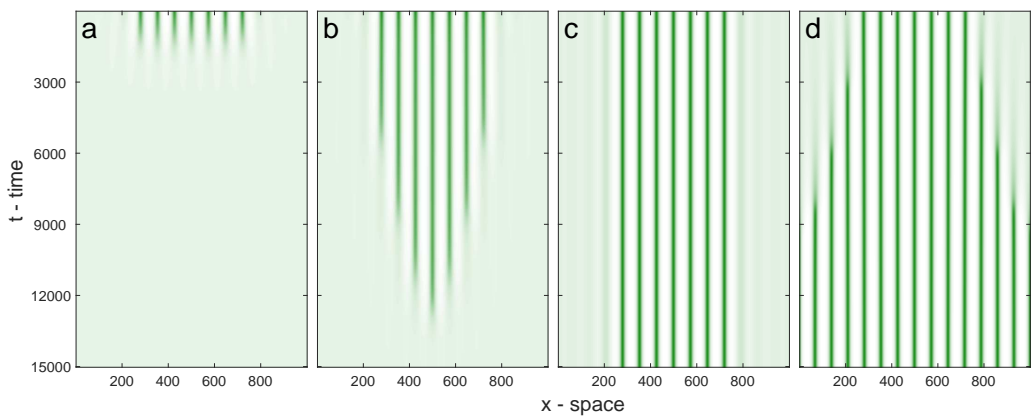}
 \caption{ \footnotesize Space-time plot showing the dynamics of the SG model with parameters (\ref{sgpar}) and $\delta_w=1200$,
 starting with an initial condition of a stable localized state of seven peaks.
  Darker green colors show denser vegetation, with the y-axis for time (going down) and the x-axis for space.
 (a) The value $p$ is decreased, so that the system is taken out of the bistability range ($p=1.006$).
 Thus, no gradual transition is possible, and an abrupt desertification shift occurs.
 (b) The value of $p$ is decreased slightly, so that the system is outside the snaking range ($p=1.0069$), but still inside the bistability range.
 Therefore a gradual desertification process takes place.
 (c) The value of $p$ is not changed, illustrating the dynamic stability of this localized state
 (d) By increasing the value of $p$ so it is outside the snaking range ($p=1.009$), but still within the bistability range, a gradual rehabilitation process takes place.
  \label{fig:SGdyn} }
\end{figure}

The dynamics for this system is symmetric, as shown in Fig. \ref{fig:SGdyn}, where we take a periodic patch from $p = 1.008$ and instantaneously change $p$ to some other values and let the system run.
In more realistic scenarios, a gradual change of $p$ might be more appropriate, and the system's behaviour will then also depend on the rate of change of $p$ \cite{Siteur2014eco_comp}.
Nevertheless, Fig. \ref{fig:SGdyn} shows that under instantaneous change of $p$ both a gradual desertification and a gradual recovery can take place,
as a transition from a localized state to either a low uniform-vegetation, or a periodic pattern, respectively.
Note that the mechanism of gradual recovery from low-biomass uniform vegetation to periodic pattern that is shown in Fig.  \ref{fig:SGdyn}d 
is different from the gradual recovery from bare soil to periodic pattern that is shown in Fig. \ref{fig:LLdyn}c. 
Rather than expanding and splitting, the patch at the fringe remains unchanged but triggers the formation of a new patch at a distance by seed dispersal (biomass diffusion). 
This difference in mechanisms may be attributed to the lower soil-water content in the uniform low-biomass areas as compared with bare-soil areas, because of water uptake, 
which prevents the expansion of the fringe patch.
We may also compare here the two possibilities of desertification dynamics, namely abrupt and gradual, by comparing Fig. \ref{fig:SGdyn}a and b (see also Fig. S2 for two dimensional systems).
In both cases we start with the same stable localized state, and end up in a uniform-state, but the transition occurs globally in the abrupt case, and via a propagation of a desertification front in the gradual case.
The abrupt transition is irreversible because of the bistability of periodic and uniform states since once a threshold is crossed the collapse is certain, 
while the gradual transition is reversible because of monostability of front solutions.
It is interesting to note that the abrupt transition may be expected to be a faster process both due to the typical physical parameters of ecosystems \cite{zelnik2015pnas},
and since in a large enough system front propagation along the whole system will always take longer than a global response across the system.

\begin{figure}[h]
 \includegraphics[width=0.755\textwidth]{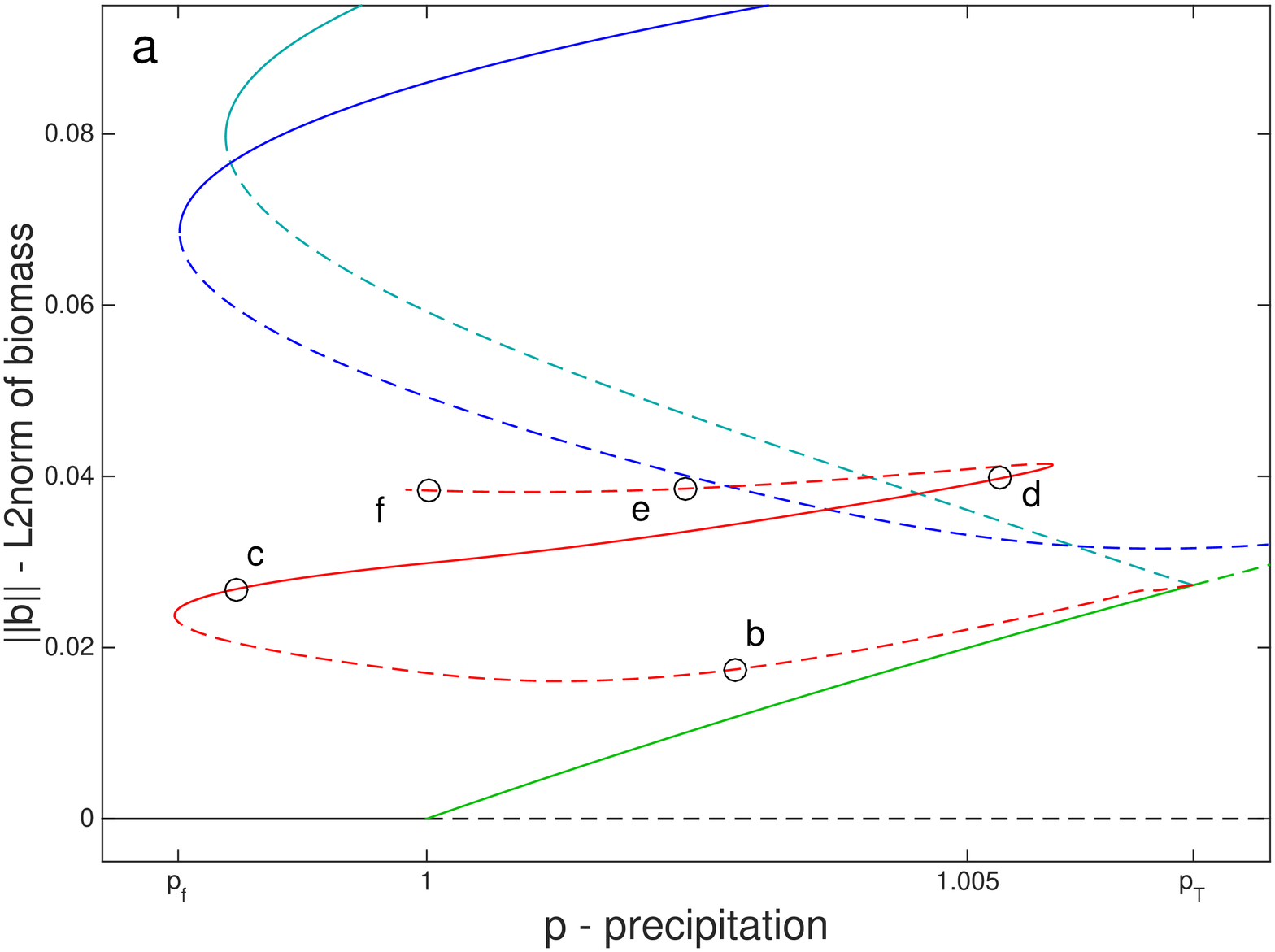}
  \includegraphics[width=0.2355\textwidth]{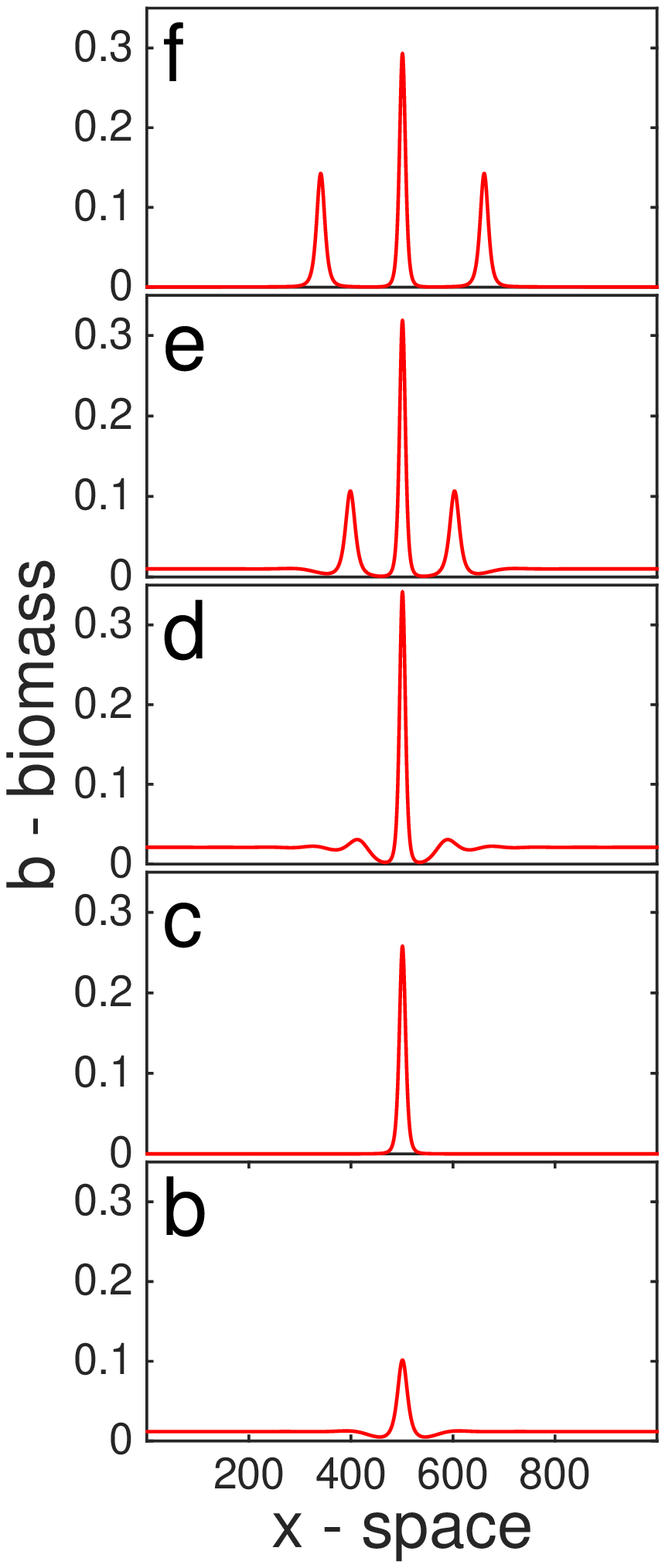}
 \caption{ \footnotesize
 (a) Bifurcation diagram of the SG model with parameters (\ref{sgpar}) and $\delta_w=1500$, showing the L2norm of the biomass versus the precipitation rate $p$.
 The black and green branches are for uniform states of bare-soil and uniform-vegetation, while the blue and cyan branches are for periodic states with different wavelengths.
 The red curve shows part of the bifurcation structure of the localized states, that fails to form a homoclinic snaking structure.
 The wavelength of the blue curve was chosen so that its left-most point is approximately at the edge of the bistability range between periodic states and bare-soil (see Fig. S1).
 Solid (dashed) lines denote stable (unstable) states.
 (b)-(f): plots of different localized states, as marked on the bifurcation diagram.
 The numerical continuation breaks down when the curve comes back to the parameter value of $p=1$ from (e) to (f), as discussed in more detail in section \ref{scoll-sec}.
 \label{fig:SGdw1500}}
 \end{figure}

By changing the model parameters, the Turing bifurcation, together with the associated branch of localized states,
can be pushed close to the bifurcation between the two uniform states of bare-soil and uniform-vegetation.
In this manner, the periodic branches which were previously bistable with the uniform-vegetation state, are now bistable with either the uniform-vegetation or bare-soil states, for different values of $p$.
This is achieved by increasing the value of $\delta_w$, the result of which is shown in the bifurcation diagram in Fig. \ref{fig:SGdw1500}a.
As shown by the red curve in the bifurcation diagram, and the corresponding system states shown in Fig. \ref{fig:SGdw1500}b-f,
a branch of localized states emanates from the vicinity of the Turing bifurcation similarly to the previous case shown in Fig. \ref{fig:SGdw1200}.
However, in our numerics the branch does not continue to snake up, but it rather stops abruptly after two folds around $p=1$, creating a Z shape.
Moreover, the edge of the new snaking range now coincides with the bistability range, similarly to the case of the LL model seen in Fig. \ref{fig:LLbif}.

Before and after the first fold point (Fig. \ref{fig:SGdw1500}b,c) the system is comprised of a single peak with either a low-biomass vegetation (panel b) or a bare-soil (panel c) background.
This type of state often occurs for a bistablity of periodic patterns and uniform states, which typically leads to either a localized states with more than one peak \cite{Bel2012theo_ecol,zelnik2015pnas},
or stays with a single peak, forming an isola-like curve \cite{Zelnik2013regime,Siteur2014eco_comp}.
As the second fold is reached (Fig. \ref{fig:SGdw1500}d) two small peaks are formed, one on either side of main peak, and these peaks slowly get larger (Fig. \ref{fig:SGdw1500}e) as we pass the second fold point.
All this takes place far from the bistability with the bare-soil, and the behavior of this branch changes when this bistability is reached at $p=1$.
As can be seen in Fig. \ref{fig:SGdw1500} between (e) and (f), the two smaller peaks slowly move away from the main peak,
and unlike the case of a typical homoclinic snaking curve, they do not stop their movement away from the center.
Shortly after (f) the numerical continuation fails, and at this point we may conclude that this is due to the translational symmetry of the system and the lack of significant interaction between the peaks,
which brings about a certain degeneracy in the solution branch: depending on fine details of the chosen algorithm (spatial discretization, numerical tolerances and continuation step size),
different situations arise such as further growth and motion of the lateral peaks, or of just one of them, with the center peak staying fixed or shrinking.

Thus, by pushing the Turing bifurcation closer to the bare-soil branch, we now have a similar behavior as in the LL model, with the same type of asymmetric dynamics (see Fig. S3):
If we take a localized state of a few patches and increase $p$, there is a finite range of $p$ where we obtain fronts of periodic vegetation domains invading the low uniform-vegetation as in Fig. \ref{fig:SGdyn}d.
On the other hand, if we decrease $p$ outside of the snaking range, for $p<p_f$ we have an abrupt transition to bare-soil, while for $p_f<p<1$ a slow rearrangement of the patches to a long wavelength periodic state will occur \cite{Zelnik2013regime}. In both cases we will not see a bare-soil domain invading the periodic vegetation one.
This contrasts with the result shown in Fig. \ref{fig:SGdyn}, where both abrupt and gradual transitions to low uniform $b$ are possible.
While the distinction between very low uniform $b$ and bare-soil $b = 0$ might at first appear overly subtle, in the remainder of this paper we discuss it from a mathematical point of view,
 and explain that the impossibility of fronts of bare-soil invading a periodic-vegetation domain is a general result for models of reaction diffusion type.

\section{Snaking Collapse} \label{scoll-sec}

We have seen two distinct types of bifurcation structures in Fig. \ref{fig:SGdw1200} and Fig. \ref{fig:SGdw1500},
which bring about different dynamical behaviors of the system, abrupt vs. gradual desertification shifts, as seen in Fig. \ref{fig:LLdyn}a and Fig. \ref{fig:SGdyn}b, respectively.
The transformation between these two types of behavior occurs as a continuous transition in the SG model following a change in parameters.
For the parameter values considered here, we find that this transition occurs for an intermediate value of water diffusion $1200<\delta_w<1500$ (see supplementary information for more details).

At this intermediate value of $\delta_w$ the fold of the single peak solution is around $p=1$, so that the fold is just about to create a bistability with the bare-soil state.
If we look more closely at the states shown in Fig. \ref{fig:SGdw1500}b-f, we can see that in the range of the stable uniform-vegetation state the tails of the peaks are oscillatory,
while they become exponential in the range of the stable bare-soil state.
Once the localized states branch has more than a single peak (Fig. \ref{fig:SGdw1500}c), it ends abruptly around $p=1$.
This signifies that with more than one peak, oscillatory tails play an important role in keeping the peaks in place, and not move away from each other.
A more thorough understanding of the tails of the localized solutions can be gained by looking at the spatial dynamics representation of the model.
For this we define $u=(b,\partial_x b,w,\partial_x w, h,\partial_x h)^T$
and transform the steady state problem written as

\begin{gather*}
0= \partial_x^2 b + f_1(b,w), \quad 0= \delta_w \partial_x^2 w + f_2(b,w,h), \quad 0= \delta_h \partial_x^2(h^2) + f_3(b,h),
\end{gather*}

which, as a second order system is reversible under $x\mapsto -x$, into the (reversible under $x\mapsto -x, u\mapsto(u_1,-u_2,u_3,-u_4,u_5,-u_6)$)
set of first order differential equations

\begin{gather} \label{sdform}
\partial_x u=G(u):=   \begin{pmatrix}
u_2 \\ -f_1(u_1,u_3) \\
u_4 \\ -\frac 1 {\delta_w} f_2(u_1,u_3,u_5) \\
u_6 \\ -\frac 1 {2\delta_h u_5} (f_3(u_1,u_5) + 2\delta_h  u_6^2 )
\end{pmatrix}.  \end{gather}
If we now treat space as time, then the uniform solutions correspond  to fixed points $u^*$ of equation (\ref{sdform}),
and the eigenvalues $\mu_j$ of the linearization $\partial_u G(u^*)$ are called their spatial eigenvalues.
Note that $u_5=h=\frac p \alpha \frac{b+q}{b+qf}>0$ for all steady states, so that $G(u^*)$ never diverges.

\begin{figure}[h]
 \includegraphics[width=0.99\textwidth]{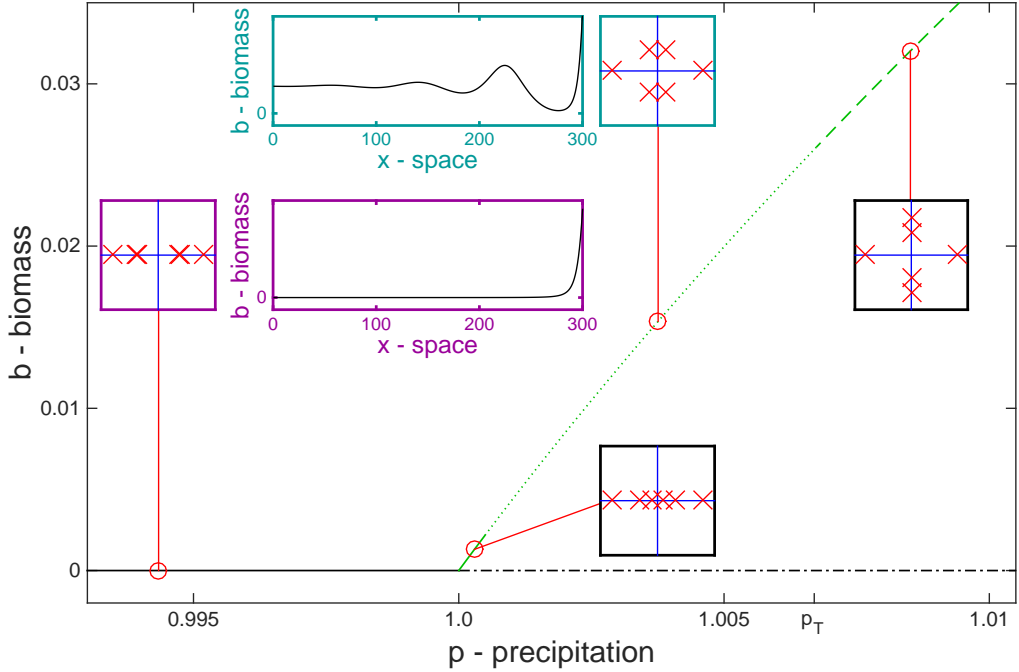}
 \caption{ \footnotesize Bifurcation diagram of the uniform branches, with information on the eigenvalues of the spatial-dynamics form of the model.
 The black and green branches are for bare-soil and uniform-vegetation states respectively.
 A solid line signifies that all six eigenvalues are real, a dotted line that there are four complex eigenvalues, while dashed lines signify that there are pure-imaginary eigenvalues.
 The square boxes show eigenvalues plotted on the complex plane, at different points along the uniform branches.
 An example of exponential and oscillatory tails is shown in the purple and cyan insets respectively, with the complementary real and complex eigenvalues in the square boxes with the same color.
 Note that the green branch is stable up to $p = p_T$ where the branch of periodic states bifurcates.
 \label{fig:SGspaEVs}}
 \end{figure}

Figure \ref{fig:SGspaEVs} shows the bifurcation diagram of the two uniform states, together with their spatial eigenvalues.
Beyond the Turing point ($p>p_T$) we have eigenvalues on the imaginary axis that relate to the periodic states in the system,
but for lower values of $p$ the four complex eigenvalues imply oscillatory tails that connect the periodic domain to the uniform-vegetation.
However, just before the uniform-vegetation branch connects with the bare-soil one all eigenvalues become real, signifying exponential tails.
For even lower values of $p$, only the bare-soil state is relevant, and it has only real-eigenvalues.
This is a prerequisite of all reaction-diffusion like models of drylands, since negative biomass values are not physical, and therefore oscillations around zero should be ruled out.

Thus, if there is a bifurcation from bare-soil to a branch of low uniform-vegetation, in a consistent model this branch generically (co--dimension one) must initially also have only real spatial eigenvalues.
In the SG model, although this initial range is small, a careful numerical analysis shows that snaking fails around the point where the eigenvalues change, and not where the uniform branches meet.
We therefore conclude that the breaking of the snaking structure occurs since without oscillatory tails the peaks can not be held in place, and instead slowly drift away.
This appears to be a case of a Belyakov--Devaney transition, see \cite[\S5.3.3]{HomSand10} for rigorous mathematical results in generic reversible ODE systems,
which essentially mean that $N$--homoclinic orbits in the ODE system (\ref{sdform}) can only be guaranteed to exist
in the parameter range where the spatial eigenvalues $\mu_j$ closest to the imaginary axis are complex, and, moreover, their period scales like $1/$Im($\mu_j$).
Note that here we refrain from analyzing the nonlinear terms, but the basic phenomena of the Belyakov--Devaney transition holds for generic reversible ODE systems,
and in this sense only depend on the linearization around the uniform steady-state.
This gives a mathematical explanation to why snaking between localized states and bare-soil does not occur in SG and related models,
even if there is a large bistability range of the periodic solutions and bare-soil, as in Fig. \ref{fig:SGdw1500}.

\section{Discussion}

Our results illuminate the complex nature of desertification in patchy ecosystems, and the important role that localized states may play in the dynamics of desertification.
Transitions (regime shifts) from uniform high-biomass vegetation to periodic vegetation, or from periodic vegetation to low-biomass vegetation, or from uniform-vegetation into bare-soil
(in an ecosystem without a stable patchy state), can occur gradually via moving fronts.
However, the transition from periodic vegetation to bare-soil is found to be an abrupt global collapse, since the absence of bare-soil fronts invading periodic patterns implies no gradual shifts to bare soil, or gradual desertification.

The absence of gradual desertification to bare-soil can be understood by its physical mechanisms as follows.
The existence of a bare-soil front invading a domain of periodic vegetation would imply a better micro environment within the domain as compared with the domain's fringe where the vegetation is dying out.
Such a front would require strong facilitation within the vegetation domain, which in the SG model can be accounted for by reduced evaporation due to shading.
Additional facilitation factors, such as enriched nutrients by litter decomposition, limited access of grazers, and reduced  wind and soil erosion, are not included in the SG model nor in other related models.
Counteracting the reduced facilitation by shading in the domain's fringe (as compared with the inner part of the domain)
 is the larger water-contributing bare-soil area that the vegetation at the domain's fringe benefits from.
That effect is strong in ecosystems showing vegetation pattern formation and wins out over the facilitation by shading (facilitation that is too strong results in uniform rather than patterned vegetation).
As a result bare-soil fronts invading periodic vegetation are not found.
For the same reason a localized single-peak solution ceases to exist in a fold bifurcation at a lower $p$ value, compared to any periodic pattern, as seen in Fig. \ref{fig:SGdw1500}.

We are not aware of empirical observations of bare-soil fronts invading vegetation patterns in flat terrains.
However, there are observations of consumer fronts, where consumers aggregate at front positions and affect their dynamics \cite{Silliman2013arees}.
Further model and empirical studies are needed to clarify the conditions under which bare-soil fronts invading periodic vegetation, and therefore gradual desertification to bare-soil, are possible.

The breakup of the snaking structure due to real eigenvalues of the spatial problem appears to be a generic behavior expected to be shared by other vegetation models.
Its occurrence due to a bare-soil state may be only one case out of many, although possibly the most interesting one from an ecological perspective.
Moreover, one may wonder if the snaking breakup has further repercussions than those described herein.
For example, in many models the wavelength of the localized states appears to change significantly within the snaking range.
In particular, models where the eigenvalues change into real values appear to have a stronger change in wavelength, implying that the eigenvalues may play a role in wavelength selection.

Since the behavior of the localized states in the SG model can be explained using spatial dynamics, it brings up the question of applying the same methodology to the LL model.
This is problematic, as rewriting the LL model into first order equations results in a term of $1/b$ in one of the equations.
Since the localized states occur in the background of bare-soil state, namely $b=0$, this term diverges, and the eigenvalues cannot be calculated.
It remains an open question how to proceed with the analysis in this case.
It is interesting to note however, that it might be this term exactly that leads to this unique situation where the whole bifurcation structure of localized states occurs in a bistability with the bare-soil state.

\section*{Acknowledgements}
We wish to thank Golan Bel, Edgar Knobloch, Jens Rademacher and Omer Tzuk for helpful discussions. We also wish to thank two anonymous reviewers for helpful comments on the manuscript.
The research leading to these results received funding from the Israel Science Foundation Grant 305/13.

\setcounter{figure}{0}
\makeatletter 
\renewcommand{\thefigure}{S\@arabic\c@figure}
\makeatother

\section{Supplementary Information}

\subsection{Continuous transition of snaking breakup}
The bifurcation diagrams in Fig. 4 and Fig. 6 show a qualitatively different structure, where the homoclinic snaking structure in Fig. 4 collapses,
and the range of gradual desertification that is between the snaking range and the region where only bare-soil is stable, disappears.
We wish to see how the transition between these two structures takes place as we change $\delta_w$, and to find how we can locate the point of transition.

In order to show that this is a smooth transition, we look at the first fold of periodic-state branches.
First, we note that the location (value of $p$) of this fold changes for different wavelengths, 
where in Fig. 4 the branch shown in blue is for the wavelength that occurs for the lowest value of $p$.
On the other hand, the first fold of the localized states occurs for a higher value of $p$, since the snaking range is well inside the bistability range.

Therefore, for the parameters of Fig. 4, if we follow the location of the fold point for different system sizes with periodic boundary conditions, 
i.e., different wavelengths, we will have a minimum for that wavelength, after which the location of the fold will increase.
This behavior can be seen in the red curve of Fig. S1, where starting from small  system size (wavelength), the position initially goes down to the minimum,
and then goes up and levels when the wavelength goes to infinity (and approaches the single peak solution).
Repeating the same procedure for the parameter values of Fig 6, no such minimum exists, as seen in the blue curve of Fig. S1.
Instead the curve goes monotonically down, and levels at a certain value (where it meets the single peak solution).
This signifies that we have a smooth transition between a snaking structure that is nested inside the bistability range, 
and the case where the edges of both the bistability range and the snaking range coincide.
Moreover, we have a well defined measure of this behavior, namely whether such a minimum occurs for a finite wavelength, 
and we can use this measure to pinpoint the parameters for which this type of bifurcation takes place.

\begin{figure}[h]
 \includegraphics[width=0.85\textwidth]{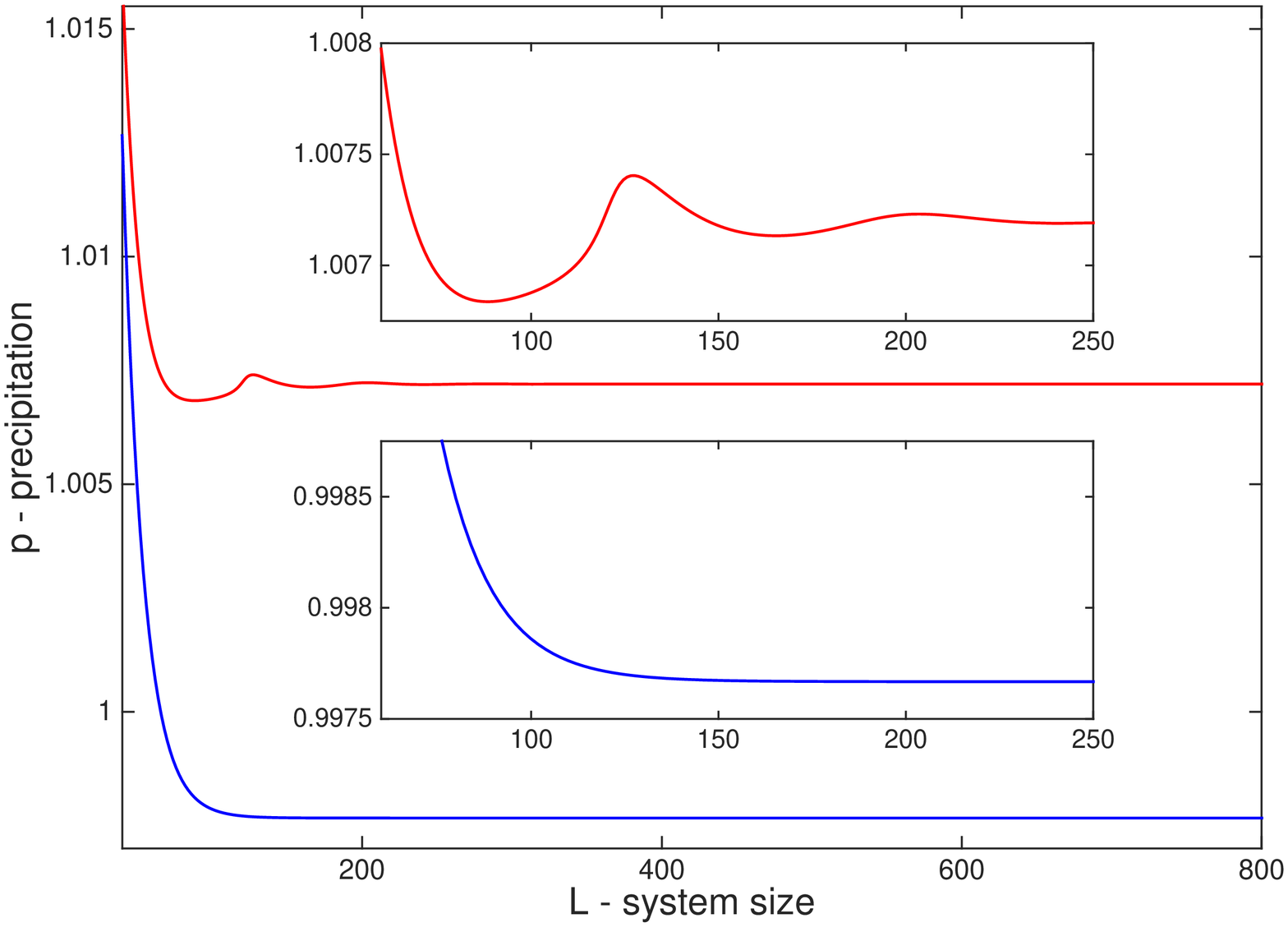}
 \caption{ \footnotesize Location of the fold point of a single peak solution, as a function of the size of the period.
 The red (blue) curve denotes that relation between the system size and value of $p$, for a value of $\delta_w$ of $1200$ ($1500$).
 \label{fig:SGfold}}
 \end{figure}

\subsection{Abrupt and gradual shifts in two dimensional systems}
The analysis done on both the LL and SG models focused on the case of a one-dimensional system, since it is much simpler to describe and present, it allows for use of spatial dynamics, and bifurcation diagrams are much easier to build. However, we believe the generality of our results also holds for the case of a two-dimensional system since:
\begin{enumerate}
\item Localized states, and their snaking range, have been found in many different models in both one and two dimensions, showing that the phenomenon is general and relevant to two dimensions as well.
\item Gradual shifts due to localized states have been found and described in several models.
\item The main inhibitor to gradual shifts and localized states that we have found, relates directly to sinusoidal tails, or lack thereof. Put simply, the fact that negative values of biomass are not physical has nothing to do with the dimension of the system, and therefore one would expect that bare-soil would not allow for gradual shifts in two dimensional systems as well. 
\end{enumerate}
In order to elucidate this generality, as well as to show how the two kinds of regime shifts look like in a two-dimensional system, we investigated the parameter space of the SG model in two spatial dimensions using time integration simulations and steady-state continuation with pde2path. We were able to find a parameter range where localized states of spots in a uniform background of low uniform vegetation exist. As shown in Fig. S2, when we decrease the value of $p$ so the system is outside the snaking range, a shift occurs. If the system is still within the bistability range of periodic states and low uniform-vegetation (Fig. S2, bottom panels), the system gradually shifts to a low uniform-vegetation state over a long period of time. If however the system is fully outside the bistability range (Fig. S2, top panels), then an abrupt and quick shift occurs to the bare-soil state.

\begin{figure}[h]
 \includegraphics[width=0.99\textwidth]{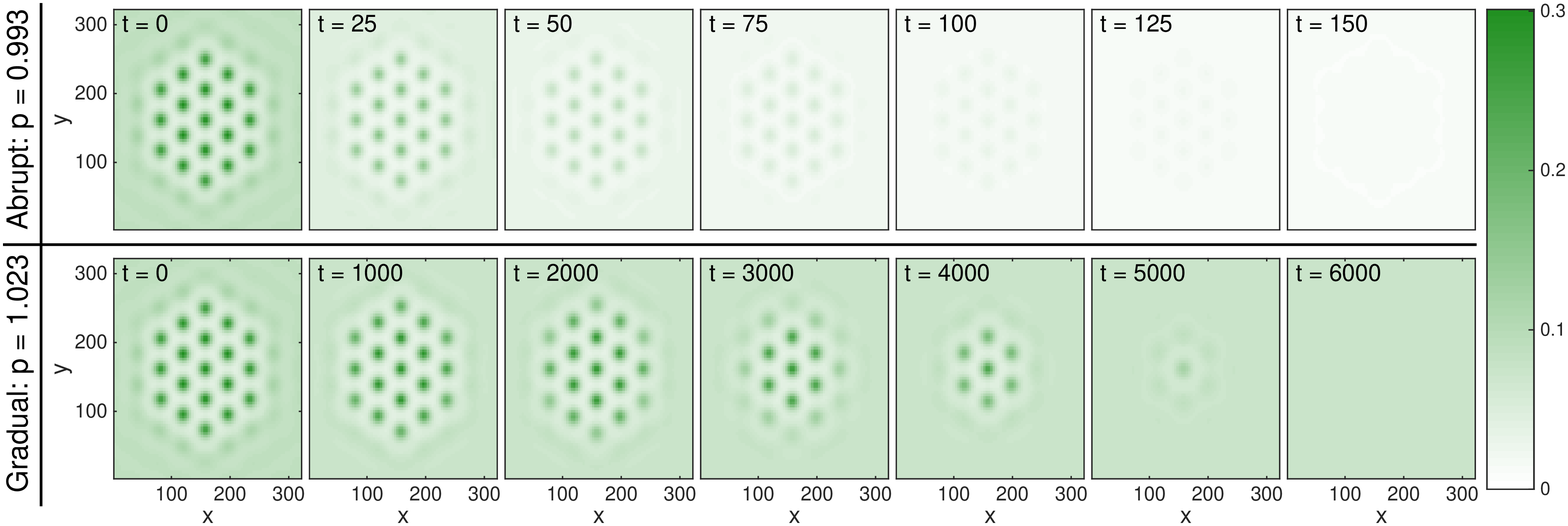}
 \caption{ \footnotesize Snapshots taken at different times of abrupt and gradual shifts of the SG model in a two-dimensional system with periodic boundary conditions. The system is initialized with a localized hexagonal-spot pattern of 19 spots at $p=1.028$, and then shifted to either $p=1.023$ for a gradual shift (bottom panels), or $p=0.993$ for an abrupt shift (top panels). Note that the time scale of these two shifts is very different, with the abrupt shift occurring much faster. The system size is 320 by 320, with $\delta_w=600$ and other parameters as in the main text. The colors depict the level of biomass $b$, and the time of each snapshot is given inside each panel.
  \label{fig:SG2d}}
 \end{figure}

\subsection{Dynamics of the SG model}
For completeness, we show in Fig. S3 the two transitions that occur for $\delta_w=1500$, when taking a localized state of seven peaks, and lowering $p$ outside of the snaking range.
In both cases we see that no gradual transition occurs into bare-soil, but rather either an abrupt transition to bare-soil, or a rearrangement of the same seven peaks into a long wavelength periodic pattern. 
We note that the specific number of peaks in the rearrangement process shown in Fig S3b does not change the qualitative dynamics, so that the eventual wavelength of the pattern is determined by initial conditions and domain size, and not by the system's parameters.

\begin{figure}[h]
 \includegraphics[width=0.9\textwidth]{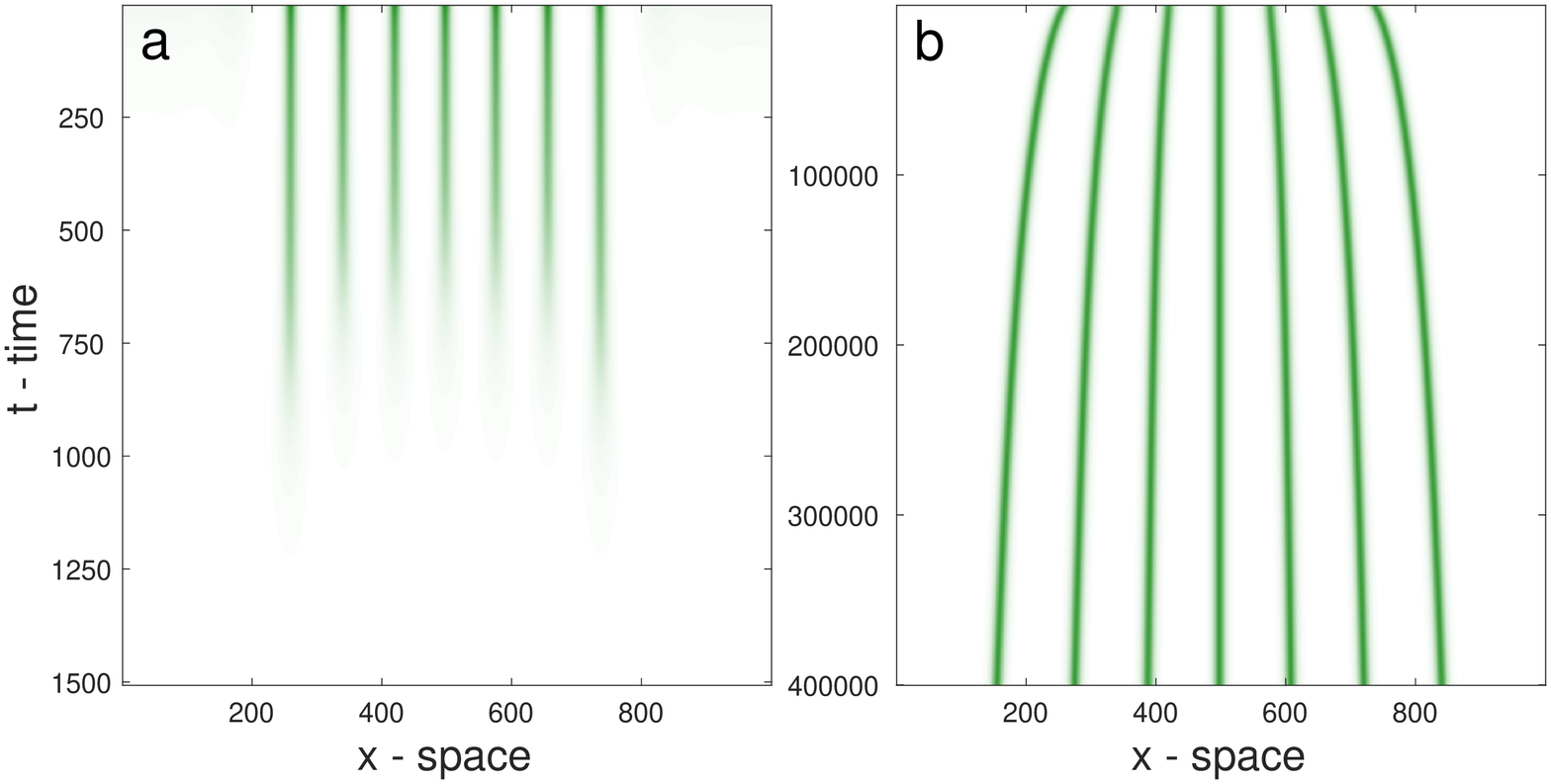}
 \caption{ \footnotesize Space-time plot showing the dynamics of the SG model with parameters of Fig. 6 ($\delta_w=1500$),
 starting with an initial condition of a stable localized state of seven peaks at $p=1.006$.
  Darker green colors show denser vegetation, with the y-axis for time (going down) and the x-axis for space.
 (a) The value $p$ is decreased to $p=0.997$, so that the system is taken out of the bistability range, bringing an abrupt collapse.
 (b) The value of $p$ is decreased enough so that a stable uniform-vegetation state does not exist ($p=0.9998$). The peaks slowly spread out in space, eventually partitioning space equally between them.
 Note the markedly different time scales used in the two panels.
  \label{fig:SGdyn1500} }
\end{figure}

We show in Fig. S4 the profiles of states of the SG model at $\delta_w=1200$, as a result of taking a localized state of seven peaks and changing the value of $p$. These correspond to horizontal cuts of Fig. 5, taken at times $t=500,2000$.

\begin{figure}[h]
 \includegraphics[width=0.9\textwidth]{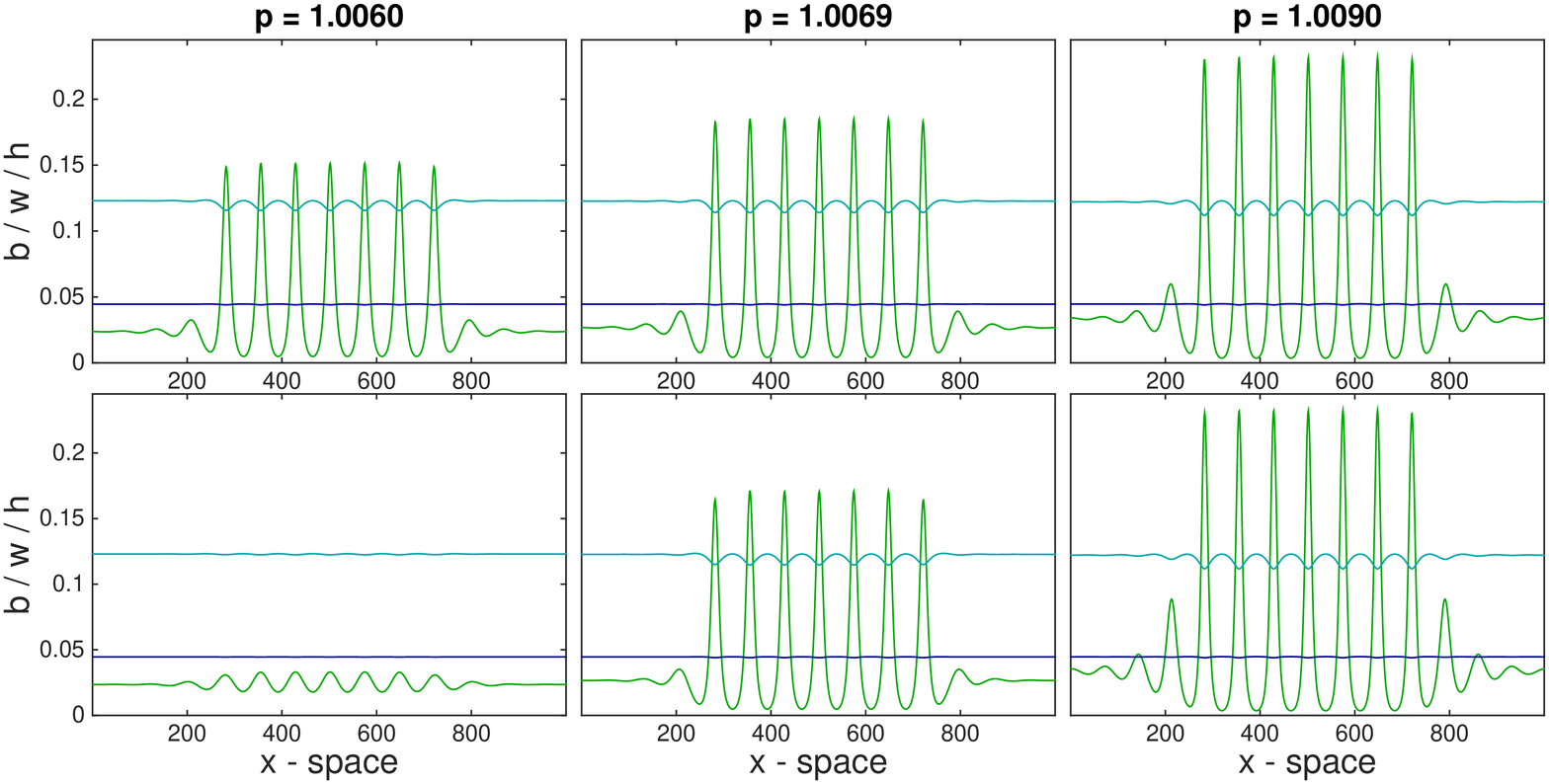} 
 \caption{ \footnotesize Profiles of states of the SG model with $\delta_w=1200$, showing the values of biomass, soil-water and surface water (green, cyan and dark-blue respectively) across space. The profiles are snapshots at time $t=500$ for the top row, and $t=2000$ for the bottom row. These correspond to snapshots of the space-time plot of Fig 5 a,b,d.
   \label{fig:SGprofiles1}}
 \end{figure}

\bibliographystyle{unsrt}
\bibliography{DFP}

\end{document}